\documentclass{article}
\usepackage{spconf, amsmath, graphicx, amssymb}
\usepackage{multirow}
\usepackage{hyperref}
\hypersetup{hidelinks,
	colorlinks=true,
	allcolors=black,
	pdfstartview=Fit,
	breaklinks=true}

\title{MEW-UNet: Multi-axis Representation Learning in Frequency Domain for Medical Image Segmentation}
\name{Jiacheng Ruan, Mingye Xie, Suncheng Xiang*, Ting Liu, Yuzhuo Fu* \thanks{* Suncheng Xiang and Yuzhuo Fu are the co-corresponding authors. This work was partially supported by the National Natural Science Foundation of China (Grant No. 61977045).}}
\address{Shanghai Jiao Tong University, Shanghai, China}
\begin{document}
%
\maketitle
\begin{abstract}
Recently, Visual Transformer (ViT) has been widely used in various fields of computer vision due to applying self-attention mechanism in the spatial domain to modeling global knowledge. Especially in medical image segmentation (MIS), many works are devoted to combining ViT and CNN, and even some works directly utilize pure ViT-based models. However, recent works improved models in the aspect of spatial domain while ignoring the importance of frequency domain information. Therefore, we propose \textbf{M}ulti-axis \textbf{E}xternal \textbf{W}eights \textbf{UNet} (\textbf{\emph{MEW-UNet}}) for MIS based on the U-shape architecture by replacing self-attention in ViT with our Multi-axis External Weights block. Specifically, our block performs a Fourier transform on the three axes of the input feature and assigns the external weight in the frequency domain, which is generated by our Weights Generator. Then, an inverse Fourier transform is performed to change the features back to the spatial domain. We evaluate our model on four datasets and achieve state-of-the-art performances. In particular, on the Synapse dataset, our method outperforms MT-UNet by 10.15mm in terms of HD95. Code is available at \emph{\href{https://github.com/JCruan519/MEW-UNet}{https://github.com/JCruan519/MEW-UNet}}.
\end{abstract}
\section{Introduction}

Medical image segmentation (MIS) can assist relevant medical staff in locating the lesion area and improve the efficiency of clinical treatment, which has great practical value. In recent years, UNet \cite{unet}, an encoder-decoder model based on U-shape architecture, has been widely utilized for MIS. Due to its strong scalability, many works are carried out based on the U-shape architecture. For example, UNet++ \cite{unet++} reduces the semantic difference between the encoder and decoder by introducing the dense connection. Att-UNet \cite{attentionunet} introduces a gating mechanism to make the model focus on the targets.


The above improvements are all based on CNNs, and the natural locality of the convolution operation makes networks obtain global information poorly. ViT \cite{vit}, due to its self-attention mechanism (SA), improves the long-range dependency modeling ability, and focuses on holistic image semantic information, which benefits intensive prediction tasks, such as image segmentation. Therefore, recent improvements can be divided into two types. On the one hand, hybrid structures based on CNNs and ViTs are widely used. For example, UCTransNet \cite{uctransnet} replaces the skip connection in UNet with the CTrans module, alleviating the problem that the skip connection between the encoder and decoder may lead to incompatible features. MT-UNet \cite{mtunet} uses CNN at a shallow level and Local-Global SA at a deep level, combined with the external attention mechanism, to obtain richer representation information. On the other hand, pure ViTs are conducted in MIS. For instance, Swin-UNet \cite{swinunet} performs better by replacing the convolution operation in UNet with the Swin Transformer Block.

Although the above models have achieved good results, they are all based on the spatial domain, and few works explore MIS in the frequency domain. The frequency domain information can help the model distinguish the lesion area and background more clearly, which is indispensable for MIS. In the general vision, GFNet \cite{gfnet} utilizes 2D discrete Fourier transform (2D DFT) to change features from the spatial domain to the frequency domain, and filters in the frequency domain are used for learning representation, which is enlightening. However, the frequency domain information is only extracted in a single axis, resulting in incomplete global information. In addition, it ignores the importance of local information in image feature extraction.

To address the problem, we propose the Multi-axis External Weights mechanism (MEW) that can simultaneously obtain more comprehensive global and local information. To be specific, the feature map is divided into four parts along the channel dimension. For the first three branches, we transform the features into the frequency domain using 2D DFT along the three different axes. Then, we multiply the frequency domain maps by the corresponding learnable weights to obtain the frequency domain information and global knowledge. In addition, depthwise separable (DW) convolution operation \cite{mobilenetv2} is conducted for the remaining branch to obtain local information. After that, via replacing SA in ViT with our MEW, Multi-axis External Weights Block (MEWB) is obtained. Finally, based on MEWB and U-shape architecture, we present MEW-UNet, a powerful network for MIS.

The contributions of this paper can be summarized as follows: 1) the Multi-Axis External Weights block is proposed to obtain global and local information at the same time, and frequency domain feature signals are introduced to help the model fully understand the context; 2) Based on U-shape and ViTs structure, the SA is replaced by our block, and a powerful MIS model, MEW-UNet is obtained; 3) We have conducted sufficient experiments on four datasets and achieved state-of-the-art results.

\section{Methods}

\subsection{Multi-axis External Weights Block}

\begin{figure}[!t]
	\begin{minipage}[b]{1.0\linewidth}
		\centering
		\centerline{\includegraphics[width=9cm]{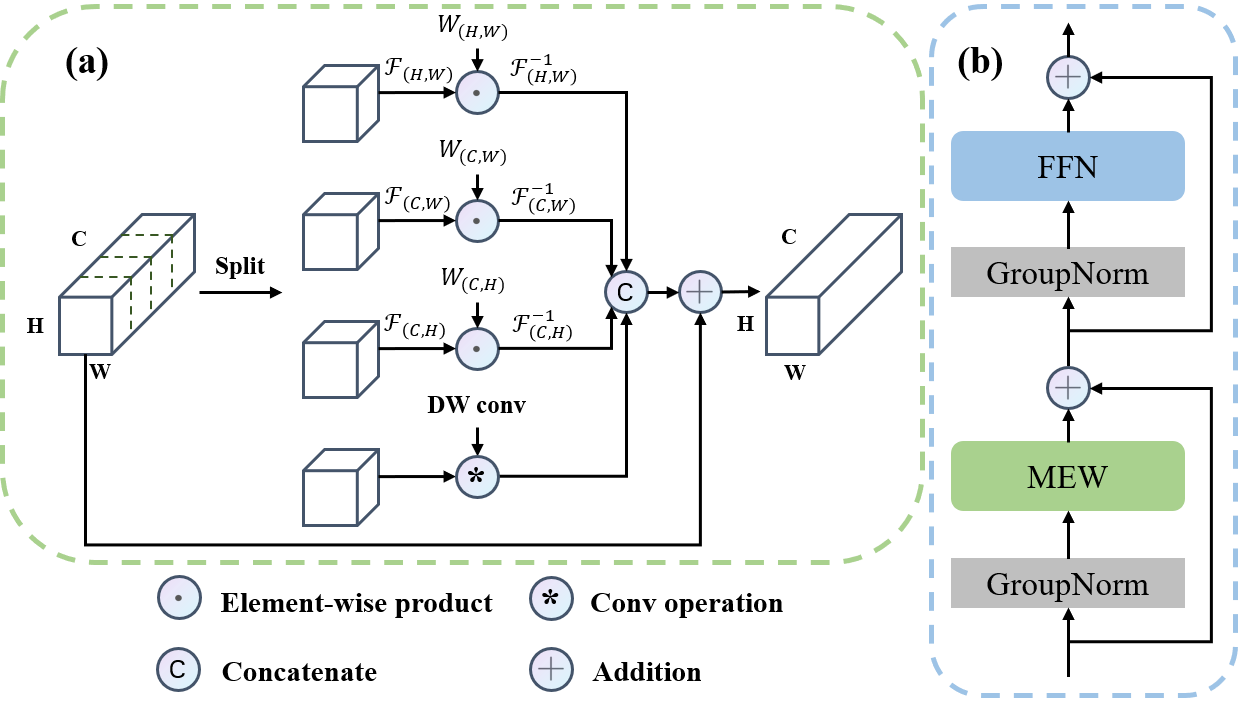}}
	\end{minipage}
	\caption{(a) Multi-axis External Weights mechanism. (b) Multi-axis External Weights block. $\mathcal{F}_{(H,W)}$ and $\mathcal{F}^{-1}_{(H,W)}$ refer to conducting 2D DFT and the 2D inverse DFT along the height and width axes of the feature map. $\mathcal{F}_{(C,W)}$, $\mathcal{F}^{-1}_{(C,W)}$, $\mathcal{F}_{(C,H)}$, and $\mathcal{F}^{-1}_{(C,H)}$ could be illustrated as the same.}
	\label{fig1}
\end{figure}

In MIS, recent methods focus more on obtaining more abundant information in the spatial domain while ignoring the importance of the frequency domain. For the spatial domain, the boundary between the segmentation object and background is often unclear, while in the frequency domain, the objects are at different frequencies, which is accessible to distinguish \cite{bibmMedicalFrequencyDomainLearning}. Although \cite{bibmMedicalFrequencyDomainLearning} and \cite{gfnet} have introduced the idea of the frequency domain, the global information they acquired is not sufficient. Therefore, we propose MEW based on 2D DFT for different axes to obtain more global information in the frequency domain, as shown in Figure \ref{fig1} (a).

Given an input feature map $X \in \mathbb{R}^{C \times H \times W}$ and $C$, $H$, $W$ is the channel, height and width of the map. Firstly, $X$ is divided into four equal parts along the channel dimension. Furthermore, we sent parts into four different branches, respectively. MEW can be expressed by formulas (1) to (4).
\begin{equation}
	x_{1}, x_{2}, x_{3}, x_{4}=\operatorname{Split}(X)
\end{equation}
\begin{equation}
	x_{i(I,J)}=W_{(I,J)} \odot \mathcal{F}_{(I,J)}[x_i]
\end{equation}
\begin{equation}
	x_{i}^\prime = \mathcal{F}^{-1}_{(I,J)}[x_{i(I,J)}], x_{4}^\prime = \operatorname{DW}(x_4)
\end{equation}
\begin{equation}
	Y = \operatorname{Concat}(x_{1}^\prime,x_{2}^\prime,x_{3}^\prime,x_{4}^\prime) + X
\end{equation}
where $i = 1,2,3$ corresponds to the first three branches. $W_{(I, J)}$ and $\mathcal{F}_{(I,J)}$ denote the learnable external weight and the 2D DFT for the corresponding axes. When $i = 1$, $I$ and $J$ represent Height-Width axes. When $i = 2$, $I$ and $J$ present Channel-Width axes. When i = 3, $I$ and $J$ represent Channel-Height axes. $\odot$ is the element-wise product. $\mathcal{F}^{-1}_{(I,J)}$ refers to the 2D inverse DFT. $\operatorname{Split}$ and $\operatorname{Concat}$ represent the split and concatenation operation along the channel dimension. 

In the first branch, we perform 2D DFT on the Height-Width axes to obtain the feature map in the frequency domain. Afterward, the element-wise product operation is performed between the feature map and the corresponding learnable external weight. Finally, we apply the 2D inverse DFT to change the map to the spatial domain. Similarly, for the second and third branches, we perform the above operations on the Channel-Width and Channel-Height axes. Thus, more abundant global information can be learned through the multi-axis operation. Note that the external weight is generated by our Weight Generator (see section 2.2). In addition, local information is equally vital in the MIS task. Therefore, for the fourth branch, we utilize the DW convolution to obtain the local information while reducing the amount of computation and parameters. Then, the feature maps of the four branches are concatenated along the channel dimension to restore the same size as the input. Finally, the residual connection of the input feature map is performed to obtain the output.

We replace our MEW with SA in ViT, as shown in Figure \ref{fig1} (b), and obtain our MEWB, which can be expressed as formulas (5) and (6).
\begin{equation}
	X' = \operatorname{MEW}(\operatorname{GroupNorm}(X)) + X
\end{equation}
\begin{equation}
	Y = \operatorname{FFN}(\operatorname{GroupNorm}(X')) + X'
\end{equation}
Note that we apply GroupNorm \cite{gn} instead of LayerNorm in vanilla ViT. Since we directly utilize $X \in \mathbb{R}^{C \times H \times W}$ as the input instead of performing the patch embedding operation. Besides, medical image dataset samples are small in size, and the batch size is often set to be small when conducting experiments. GroupNorm with four groups is conducted in our network to reduce the impact of batch size.

\subsection{External Weights Generator}

MIS belongs to the layout-specific task. Namely, the variance between samples is slight, but within samples is significant in a certain medical dataset. Only using the learnable weights initialized randomly can not learn the semantic relationship between different regions well \cite{tvconv}. Therefore, this paper introduces an External Weights Generator to transform the initialized weights. 

Specifically, the External Weights Generator is composed of several DW convolutions. For the weight corresponding to the Height-Width axes, we initialize a learnable tensor followed by a bilinear interpolation to make its dimensions the same as the input. Then, three 2D Inverted residual blocks \cite{mobilenetv2} are performed on the learnable weight. We apply the same operations as before for the weights of the Channel-Width and Channel-Height axes, except for replacing 2D Inverted residual blocks with 1D Inverted residual blocks.


\subsection{MEW-UNet}

\begin{figure}[!t]
	\begin{minipage}[b]{1.0\linewidth}
		\centering
		\centerline{\includegraphics[width=8cm]{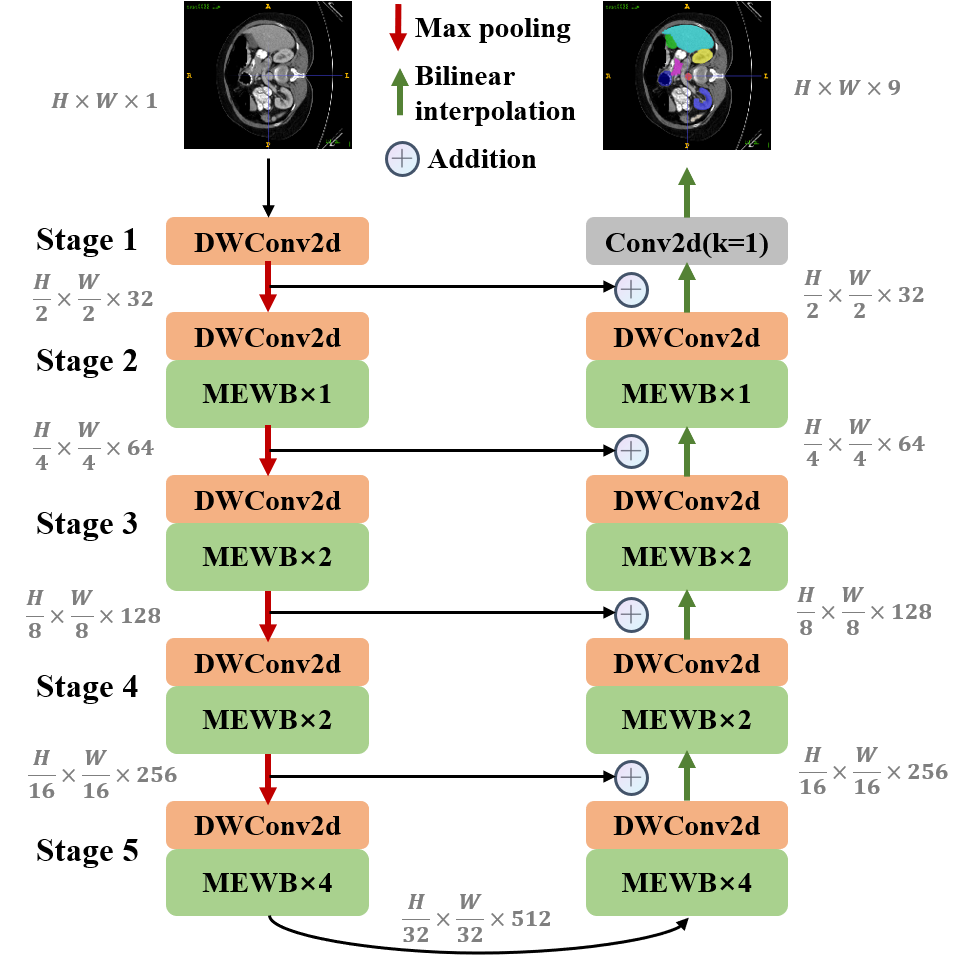}}
	\end{minipage}
	\caption{The illustration of our MEW-UNet.}
	\label{fig2}
\end{figure}

The structure of our MEW-UNet is shown in Figure \ref{fig2}. This model is based on the five-stage U-shape architecture and consists of symmetric encoders and decoders. In the first stage, only the DW convolution is used to aggregate the local features. In the following four stages, DW convolution operation is first used to change the number of channels, and then the MEWB is applied to obtain the global and local information. In addition, the number of channels is \{32, 64, 128, 256, 512\}, and the number of MEWB used in the last four stages is \{1, 2, 2, 4\}. It is worth noting that our model does not need to be pre-trained, unlike previous models, such as TransUNet and TransFuse.

\begin{table}[!t]
	\setlength\tabcolsep{2.5pt}
	\renewcommand\arraystretch{1.25}
	\small
	\caption{Comparative experimental results on the ISIC17 and ISIC18 dataset. (\textbf{Bold} indicates the best.)}
	\begin{center}
		\begin{tabular}{c|c|ccccc}
			\hline
			\textbf{Dataset} &\textbf{Model}          & \textbf{mIoU}  & \textbf{DSC}   & \textbf{Acc}   & \textbf{Spe}   & \textbf{Sen}   \\ \hline
			\multirow{4}{*}{ISIC17} &UNet \cite{unet}                   & 76.98          & 86.99          & 95.65          & 97.43          & 86.82          \\
			&UTNetV2 \cite{utnetv2}                & 77.35          & 87.23          & 95.84          & \textbf{98.05}          & 84.85          \\
			&TransFuse \cite{transfuse}              & 79.21          & 88.40          & 96.17          & 97.98          & 87.14 \\
			&\textbf{MEW-UNet (Ours)} & \textbf{81.38} & \textbf{89.73} & \textbf{96.57} & 97.99 & \textbf{89.52}        \\ \hline
			\multirow{7}{*}{ISIC18}&UNet \cite{unet}                    & 77.86          & 87.55          & 94.05          & 96.69          & 85.86          \\
			&UNet++ \cite{unet++}                 & 78.31          & 87.83          & 94.02          & 95.75          & 88.65          \\
			&Att-UNet \cite{attentionunet}               & 78.43          & 87.91          & 94.13          & 96.23          & 87.60          \\
			&UTNetV2 \cite{utnetv2}                & 78.97          & 88.25          & 94.32          & 96.48          & 87.60          \\
			&SANet \cite{sanet}                  & 79.52          & 88.59          & 94.39          & 95.97          & 89.46          \\
			&TransFuse \cite{transfuse}              & 80.63          & 89.27          & 94.66          & 95.74          & \textbf{91.28} \\
			&\textbf{MEW-UNet (Ours)} & \textbf{81.90} & \textbf{90.05} & \textbf{95.18} & \textbf{96.99} & 89.57          \\ \hline
		\end{tabular}
		\label{tab1}
	\end{center}
\end{table}

\begin{table*}[!t]
	\setlength\tabcolsep{1.5pt}
	\renewcommand\arraystretch{1.25}
	\small
	\caption{Comparative experimental results on the Synapse dataset. DSC of every single class is also reported.}
	\begin{center}
		\begin{tabular}{c|cc|cccccccc}
			\hline
			\textbf{Model}          & \textbf{DSC(\%)$\uparrow$} & \textbf{HD95(mm)$\downarrow$} & \textbf{Aorta} & \textbf{Gallbladder} & \textbf{Kidney(L)} & \textbf{Kidney(R)} & \textbf{Liver} & \textbf{Pancreas} & \textbf{Spleen} & \textbf{Stomach} \\ \hline
			V-Net \cite{vnet}                   & 68.81            & -                 & 75.34          & 51.87                & 77.10              & \textbf{80.75}     & 87.84          & 40.05             & 80.56           & 56.98            \\
			UNet \cite{unet}                  & 76.85            & 39.70             & 89.07 & \textbf{69.72}       & 77.77              & 68.60              & 93.43          & 53.98             & 86.67           & 75.58            \\
			Att-UNet \cite{attentionunet}                & 77.77            & 36.02             & \textbf{89.55}          & 68.88                & 77.98              & 71.11              & 93.57          & 58.04             & 87.30           & 75.75            \\
			TransUnet \cite{transunet}              & 77.48            & 31.69             & 87.23          & 63.13                & 81.87              & 77.02              & \textbf{94.08}        & 55.86             & 85.08           & 75.62            \\
			UCTransNet \cite{uctransnet}             & 78.23            & 26.75             & -              & -                    & -                  & -                  & -              & -                 & -               & -                \\
			MT-UNet  \cite{mtunet}               & 78.59            & 26.59             & 87.92          & 64.99                & 81.47              & 77.29              & 93.06          & \textbf{59.46}    & 87.75           & \textbf{76.81}           \\ \hline
			\textbf{MEW-UNet (Ours)} & \textbf{78.92}   & \textbf{16.44}    & 86.68          & 65.32                & \textbf{82.87}             & 80.02              & 93.63          & 58.36             & \textbf{90.19}  & 74.26          \\ \hline 
		\end{tabular}
		\label{tab2}
	\end{center}
\end{table*}

\begin{table}[!t]
	\setlength\tabcolsep{2.5pt}
	\renewcommand\arraystretch{1.2}
	\caption{Comparative experimental results on the ACDC dataset. DSC of every single class is also reported.}
	\begin{center}
		\begin{tabular}{c|c|ccc}
			\hline
			\textbf{Model}           & \textbf{DSC(\%)} & \textbf{RV}    & \textbf{Myo}   & \textbf{LV}    \\ \hline
			R50 UNet \cite{transunet}                & 87.60            & 84.62          & 84.52          & 93.68          \\
			R50 AttnUNet \cite{transunet}             & 86.90            & 83.27          & 84.33          & 93.53          \\
			R50 ViT \cite{transunet}                  & 86.19            & 82.51          & 83.01          & 93.05          \\
			TransUNet \cite{transunet}                & 89.71            & 86.67          & 87.27          & 95.18          \\
			Swin-Unet \cite{swinunet}               & 88.07            & 85.77          & 84.42          & 94.03          \\
			MT-UNet  \cite{mtunet}                & 90.43            & 86.64          & \textbf{89.04} & \textbf{95.62} \\ \hline
			\textbf{MEW-UNet (Ours)} & \textbf{91.00}   & \textbf{88.82} & 88.61          & 95.56          \\ \hline
		\end{tabular}
		\label{tab3}
	\end{center}
\end{table}

\begin{table}[!t]
	\setlength\tabcolsep{2pt}
	\renewcommand\arraystretch{1}
	\caption{Ablation Studies for our MEWB on the ISIC18 dataset. \textbf{$\circ$} denotes replacing GroupNorm with BatchNorm.}
	\begin{center}
		\begin{tabular}{cccc|cc}
			\hline
			DW & \textbf{$W_{(H, W)}$} & \textbf{$W_{(C, W)}$} & \textbf{$W_{(C, H)}$} & \textbf{mIoU} & \textbf{DSC} \\ \hline
			\textbf{$\circ$}  &                       &                       &                       &     80.06          &      88.92        \\
			\textbf{\checkmark}  &                       &                       &                       &     80.14          &      88.98        \\
			\textbf{\checkmark}  & \textbf{\checkmark}            &                       &                       &  80.82         &      89.39    \\
			\textbf{\checkmark}  & \textbf{\checkmark}            & \textbf{\checkmark}            &                       &    81.27           &     89.46         \\
			& \textbf{\checkmark}            & \textbf{\checkmark}            & \textbf{\checkmark}            &    81.29           &     89.68         \\ 
			\textbf{\checkmark}  & \textbf{\checkmark}            & \textbf{\checkmark}            & \textbf{\checkmark}            &    81.90           &     90.05         \\ 
			\hline
		\end{tabular}
		\label{tab4}
	\end{center}
\end{table}

\section{EXPERIMENTS}

\subsection{Datasets and Evaluation}

\textbf{ISIC17} and \textbf{ISIC18}, two public datasets of skin lesion segmentation, have 2150 and 2694 dermoscopy images with segmentation mask labels, respectively. We randomly divide datasets in a ratio of 7:3 for training and testing. We report five evaluation metrics, including Mean Intersection over Union (mIoU), Dice Similarity Coefficient (DSC), Accuracy (Acc), Sensitivity (Sen), and Specificity (Spe).

\textbf{Synapse}, a public multi-organ segmentation dataset, has 30 abdominal CT cases, including abdominal organs (aorta, gallbladder, spleen, left kidney, right kidney, liver, pancreas, spleen, stomach). Following \cite{transunet}, we use 18 cases for training and 12 cases for testing. Dice Similarity Coefficient (DSC) and 95\% Hausdorff Distance (HD95) are used to evaluate our method on this dataset.

\textbf{ACDC}, a public cardiac MRI dataset, is composed of 100 MRI scans with three organs, including left ventricle (LV), right ventricle (RV) and myocardium (MYO). Following \cite{mtunet}, we use the same split of 70 training cases, 10 validation cases and 20 testing cases. Dice Similarity Coefficient (DSC) is reported as the metric.

\subsection{Implementation details}

Empirically, for the ISIC17 and ISIC18 datasets, all images are resized to 256 × 256. For the Synapse dataset, all images are resized to 224 × 224. To avoid overfitting, data augmentation is performed, including random flip and random rotation. The loss function is BceDice loss. We set batch size equal to 8, and utilize CosineAnnealingLR as the scheduler. Here, we itemize the initial learning rate (lr), maximum epochs (ep), and optimizer (opt) for three datasets:
\begin{itemize}
	\setlength{\itemsep}{0pt}
	\setlength{\parsep}{0pt}
	\setlength{\parskip}{0pt}
	\item ISIC17 and ISIC18: lr=1e-3; ep=300; opt=AdamW.
	\item Synapse: lr=3e-3; ep=600; opt=SGD. 
\end{itemize}

Besides, the same settings as \cite{mtunet} are conducted for the ACDC dataset, except that we set batch size equal to 4.

\subsection{Comparison with State-of-the-Arts}

\begin{figure}[!t]
	\begin{minipage}[b]{1.0\linewidth}
		\centering
		\centerline{\includegraphics[width=7cm]{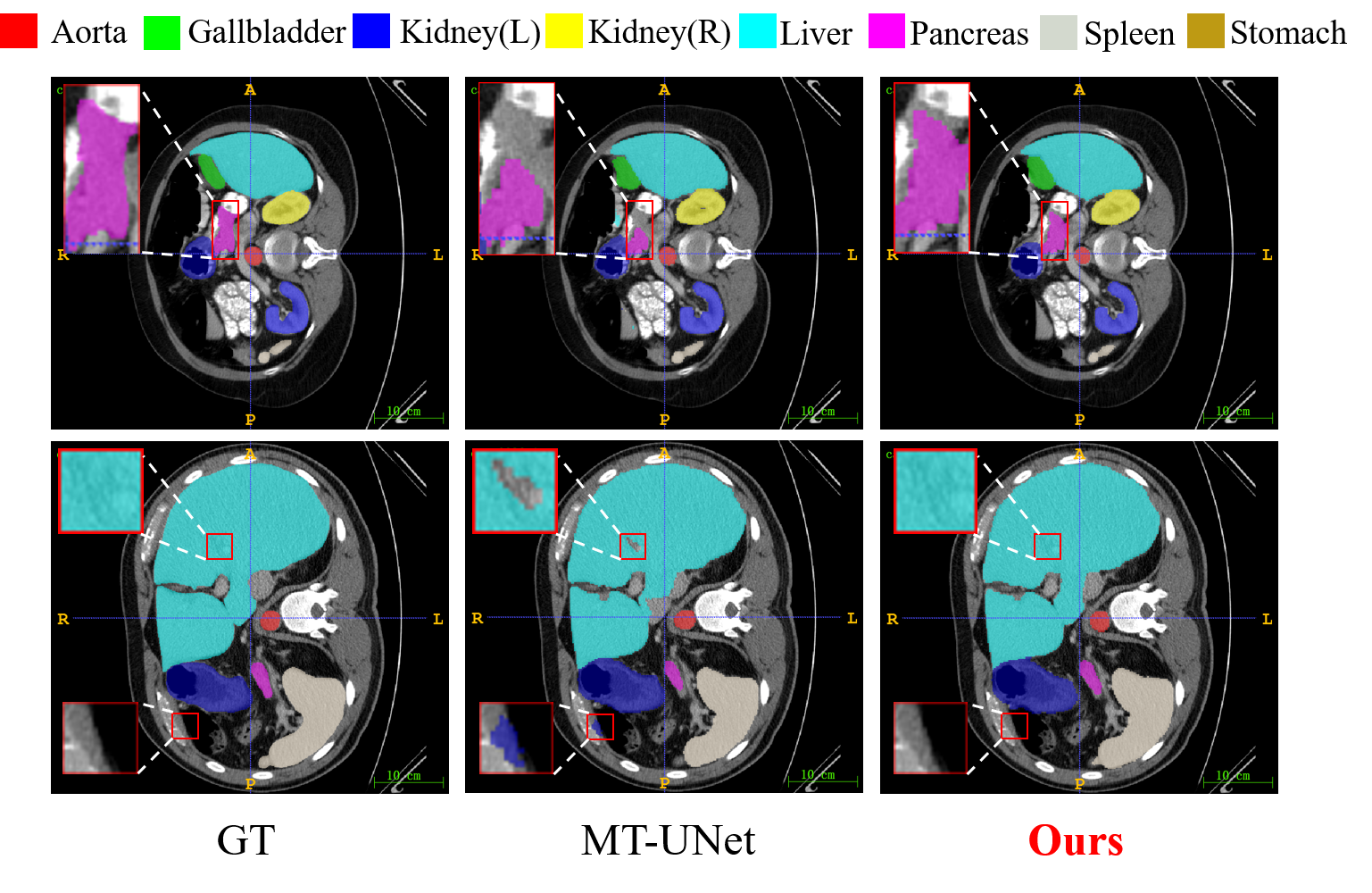}}
	\end{minipage}
	\caption{The visualization comparison on the Synapse dataset.}
	\label{fig3}
\end{figure}

We compare our model with State-of-the-Arts in recent years, such as MT-UNet (\emph{ICASSP22}) \cite{mtunet}, UCTransNet (\emph{AAAI22}) \cite{uctransnet}, TransFuse (\emph{MICCAI21}) \cite{transfuse}, SANet (\emph{MICCAI21}) \cite{sanet}, and so on. The experimental results are shown in Table \ref{tab1}, Table \ref{tab2} and Table \ref{tab3}. For the ISIC17 and ISIC18 datasets, our MEW-UNet outperforms all other state-of-the-arts on the mIoU and DSC metrics. For the Synapse dataset, our MEW-UNet is \textbf{0.33\%} and \textbf{0.69\%} higher than MT-UNet and UCTransNet in terms of DSC. Besides, it is worth noting that our model surpasses MT-UNet and UCTransNet \textbf{10.15mm} and \textbf{10.31mm} in the aspect of HD95. For the ACDC dataset, our MEW-UNet also achieves the best DSC metric. Some segmentation results are visualized by ITK-SNAP \cite{ITK-SNAP} in Figure \ref{fig3}.

\subsection{Ablation Studies}

The core idea of our MEW-UNet is multi-axis operations. Thus, we increase the number of multi-axis operations one by one to verify the effectiveness of our proposed block. The ablation performances are shown in Table \ref{tab4}, which reveals that our multi-axis operations can help in modeling more comprehensive frequency domain information and global knowledge. In addition, we also replace GroupNorm with BatchNorm to prove that the former is a better choice.

\section{CONCLUSIONS}

In this paper, we propose a Multi-axis External Weights block for acquiring frequency domain information and global knowledge. And our Weight Generator is applied to obtain external weight for medical image segmentation. Furthermore, we replace the self-attention mechanism with our block to construct MEW-UNet. Experiments show that our model achieves state-of-the-art performances. It is believed that our work will lead to new enlightenment for the subsequent model developments in the frequency domain.


\bibliographystyle{IEEEbib}
\bibliography{strings,refs}

\end{document}